\documentclass[runningheads]{llncs}

\PassOptionsToPackage{hyphens}{url}
\usepackage{hyperref}
\usepackage{mwe}
\usepackage{algorithm}
\usepackage{algpseudocode}
\usepackage{amsmath}
\usepackage{amsfonts}

\usepackage{amsthm}
\usepackage{enumitem}
\usepackage{pifont}
\usepackage{subfig}
\usepackage{graphics}
\usepackage[space]{grffile}
\usepackage{graphicx}
\usepackage{caption}
\usepackage[noabbrev]{cleveref}
\usepackage{multirow}
\usepackage{makecell}
\usepackage{tcolorbox}
\usepackage{todonotes}
\usepackage{bigstrut}
\usepackage{sepnum}

\newcommand\blfootnote[1]{%
	\begingroup
	\renewcommand\thefootnote{}\footnote{#1}%
	\addtocounter{footnote}{-1}%
	\endgroup
}

\makeatletter

\makeatletter
\crefformat{section}{\S#2#1#3}
\crefformat{subsection}{\S#2#1#3}
\crefformat{subsubsection}{\S#2#1#3}

\newcommand{\nnum}[1]{\sepnum{.}{,}{}{#1}}

\tcbset{boxsep=1mm,top=1mm,bottom=1mm,boxrule=0.1mm,enlarge top by=2mm}

\begin{document}
\sloppy

\author{
	Konstantina Dritsa* \and
	Thodoris Sotiropoulos* \and
	Haris Skarpetis \and
	Panos Louridas}
\authorrunning{K. Dritsa, T. Sotiropoulos et al.}

\institute{Athens University of Economics and Business, Athens, Greece\\
	\email { \{dritsakon, theosotr, p3110180, louridas\}@aueb.gr} 
	\medskip
	\newline{* These authors contributed equally to this work.}}

\title{Search Engine Similarity Analysis: A Combined Content and Rankings Approach}
\maketitle

\begin{abstract}
	
	\vspace{-3mm}
	How different are search engines? The search engine wars are a favorite
	topic of on-line analysts, as two of the biggest companies in the
	world, Google and Microsoft, battle for prevalence of the web search
	space. Differences in search engine popularity can be explained by
	their effectiveness or other factors, such as familiarity with the
	most popular first engine, peer imitation, or force of habit. 
	In this work we present a thorough analysis of the affinity
	of the two major search engines, Google and Bing, along
	with DuckDuckGo, which goes to great lengths to emphasize
	its privacy-friendly credentials. To do so, we collected
	search results using a comprehensive set of 300 unique queries
	for two time periods in 2016 and 2019, and developed a new
	similarity metric that leverages both the content and the
	ranking of search responses. We evaluated the characteristics
	of the metric against other metrics and approaches
	that have been proposed in the literature, and used
	it to (1) investigate the similarities of search engine results,
	(2) the evolution of their affinity over time,
	(3) what aspects of the results influence similarity,
	and (4) how the metric differs over different kinds of search services.
	We found that Google stands apart, but Bing and DuckDuckGo are largely
	indistinguishable from each other.
	\blfootnote{The final authenticated version is available online
		at \url{https://doi.org/10.1007/978-3-030-62008-0_2}.}

\keywords{search engines \and distance metrics \and results ranking \and document
	similarity}

\end{abstract}

\section{Introduction}
Search engine battles make headlines in the international media;
changes in their algorithms, aiming to produce more accurate results
that respond better to user needs, have become topics of on-line
analysts, while their rollout is eagerly followed across the globe.

Meanwhile, the inner workings of search engines are corporate
secrets. For example, although we know that Google started with
PageRank, we also know that the current Google service uses lots of
proprietary mechanisms that have not been made public. That is also
true for its prime competitor, Bing, or other search engines with a
considerable following, like Baidu.

The battle for prevalence in the search engine market is an ongoing
game. Recent developments, such as the disclosures of personal data
misuse and the advent of stricter data protection policies, affect the
dynamics of the market. A deeper look in the United States search
engine market developments over the last three years reveals that
Google's market share increased by 5.45\%, Bing's market share
decreased by 18.13\%, while DuckDuckGo's market share rose almost by a
factor of four~\cite{stat}. The latter is a search engine that goes at
lengths to satisfy privacy conscious users. It gets its results from
over four hundred sources like Wikipedia and a variety of partners
including Bing~\cite{DDGsources}. Although it has a very small share
compared to its aforementioned competitors, its considerable
increase indicates an upward dynamic against the established and
mature competitors of the market.

At the same time, search engines are evolving at a fast pace,
returning far richer results than the ``ten blue links'' of the
past~\cite{wang2018whole}. Nowadays, search
results include elaborate page titles and detailed
textual snippets, while also aggregating relevant
content from multiple specialized search services
(referred to as verticals),
such as images, video, business maps or weather
forecasts~\cite{arguello2011aggregate,wang2018whole,arguello2009vertical}.
These advances have given prominence to new user interaction
patterns, e.g., users are usually attracted
by vertical results which in turn increases the probability
of inspecting nearby web results~\cite{danqi2012blue,zeyang2015influence}.
As a result, even though the conventional approach
of result ranking can still be used for search engine
comparisons, it is essentially a first-order approximation
of the problem that does not take
into account the current heterogeneous user experience. 

In this work we tackle the question of the similarity of search
engines, by investigating whether their search results are indeed
different. We examine Google and Bing, together with DuckDuckGo.
Existing studies on search engine similarity primarily focus only on
the ranking of the search results, while not considering the rest of
their features, such as snippets or titles that play a critical role
on attracting users' clicks
\cite{clarke2007influence,joachims2007evaluating,lewandowski2008retrieval,cutrell2007you}.
Therefore, to compare these three search engines we propose a new
similarity metric that takes into account both the top $k$ lists of
search results, and their semantic content, as shown by the titles and
text snippets in their responses. We apply our metric to data from a
comprehensive set of queries gathered from two periods, in~2016
and in~2019.

\paragraph{ Contributions}
Our work contributes to both search engine affinity analysis and the
top $k$ results literature:
\begin{itemize}
\item \textbf{ Search engine affinity:} We develop an experimental 
setting
  for assessing the affinity of three search engines,
  namely Google, Bing and DuckDuckGo. By assembling a varied
  set of around 300 unique queries and inspecting
  their top 10 results over two distinct periods,
  in 2016 and 2019,
  we study and compare the
  behavior of different search engines across
  time.
\item \textbf{A novel metric for search engine similarity that
    considers both the order and the content of the results:}
Existing search engine comparison methods leverage
ranked results. However, search engines tend to 
  return rich and heterogeneous
  forms of information
  (such as elaborate titles
  and detailed snippets),
  and thus transforming
  user interaction patterns. We therefore
  introduce a combined content and rankings
  approach that returns more expressive
  similarity scores and distinguishes important
  differences in search engine behavior that are
  not apparent with existing rank-distance metrics.
\item \textbf{Comparison findings:} While Google appears to be
different than both Bing and
DuckDuckGo, the last two are indistinguishable from each other. 
\end{itemize}

We find that while Google appears to be different than both Bing and
DuckDuckGo, the last two are indistinguishable from each other. To arrive
at this result, we start in Section~\ref{sec:background} with an
overview of related work. We introduce our metric in
Section~\ref{sec:metric} and its application on our data set in
Section~\ref{sec:engine-comparison}. We wrap up with our conclusions
and discussion in Section~\ref{sec:conclusions-discussion}.

\section{Background and Related Work}
\label{sec:background}


The issues of affinity, performance, and stability in search engines
have attracted research attention since their early days in the
1990s.
The oldest studies~\cite{gordon1999finding,bar1999search,rosenthal1999search,marchionini1996search,srivastava1999precision}
focused mainly on evaluating and comparing the performance of search
engines, employing usually a few queries (2 to~20) and manually
examining the retrieved results for relevance with each search query.
None of the search engines examined in these studies survives to
this day.
In 2004, three search engines, Google and two defunct ones, were
evaluated, with Google demonstrating the best
performance~\cite{vaughan2004new}.

Concerning the affinity of search results, early studies showed that
search engines produced mostly unique results with low
overlap~\cite{etzioni1995metacrawler,marchionini1996search,krishna1998overlap}. This could be justified by the fact that each
search engine would index less than than one-third of the indexable
web~\cite{lawrence1998searchwww}, leading half of all users to try
different search engines when looking for an answer to a specific
query~\cite{white2009search}. The trend of low similarity continued
throughout the 2000s~\cite{spink2006study,bar2006methods}. A study in
2007 revealed
that the ranking of the search results by search engines was
different than how users would rank them, thus demonstrating the potential
for personalized search~\cite{bar2007user}.

In 2010, Zaragoza et al.~\cite{zaragoza2010web} conducted an
alternative approach with quantitative statements, which they
tested with 1000 queries on Google, Microsoft Live Search, and Yahoo!
Search. They found that the three search engines gave satisfactory
results for navigational queries (i.e., queries that referred to a
particular web page or service) and for frequent non-navigational
queries.

At the same time, Webber et al. ~\cite{webber2010similarity} developed
Rank-Biased Overlap, a similarity metric for ranked lists. The
researchers created a set of 113 queries and inspected the top-100
{\sc url}s produced by 11 search engines. Google and Microsoft Live
Search results were common by 25\%. Moreover, when checking against
the localized versions of the search engines (e.g., the .au domain),
Google was found to use less localization than Yahoo and Microsoft
Live.

In a subsequent work in 2011 that investigated the ranking similarity
between Bing and Google~\cite{cardoso2011google}, Cardoso and
Magalh\~{a}es applied the Rank-Biased Overlap on the results of
40,000 queries, showing that the search engines differed considerably.
Furthermore, they looked into the diversity of search results for a
given query using the Jensen-Shannon divergence and came to the
conclusion that Bing tended to interpret a given query more diversely
than Google.

In 2014, Collier and Konagurthu~\cite{collier2014information} proposed
a measure for the comparison of two top $k$ lists, based on the
minimum length encoding framework developed by
Wallace~\cite{wallace2005statistical}. The researchers used 250
queries to retrieve search results from Ask, Google, and Yahoo. They
measured the similarity of these search engines for up to the top 100
results and found that the search engines results differed linearly on
their ranks, or quadratically using the Spearman and Kendall
distances.
Agrawal et al.~\cite{agrawal2015study} proposed two methods,
TensorCompare and CrossLearnCompare, to compare search engine
affinity based on the content of web results, and used
these techniques to compare Google and Bing. We will return
to this study in~\ref{ssec:other-approaches}, where we use it to
validate our own approach.



Most of the similarity studies focus only on the ranking
of the search results, while not considering
the rest of their semantic features, such as snippets or titles.
The idea of a combined approach that facilitates both
ranking and textual semantics has appeared 
in the field of Web search, especially for
ranking search results~\cite{bian2010query,zhuang2008semantic}.
Although semantic features are largely incorporated into
the process of producing and ranking search results,
they are not integrated in the commonly used rank-distance
metrics for comparing
results between different search engines.
Moreover, in cases where the results of the search engines
have very similar {\sc url}s and rankings but very different
snippets / titles differ, a rank-distance metric cannot reflect
this dissimilarity. On the other hand,
approaches that solely focus on the content of web results
would not sufficiently represent reality,
when comparing search engines
with similar snippets / titles,
but different rankings. 

In addition, the trend towards aggregation
of multiple information sources into search
results~\cite{arguello2011aggregate,zeyang2015influence,wang2018whole,arguello2009vertical}
has significantly transformed the landscape of search result
pages, which in turn has resulted in changes in
the corresponding evaluation methodologies~\cite{bailey2010whole,wang2018whole}.
As more heterogeneous information appears on modern
search engine result pages, studies highlight
interesting user interaction
patterns~\cite{danqi2012blue,lagun2014attention,zeyang2015influence,yue2010bias}
where the sequential order of results does not play
the sole role in browsing result pages. Recent research has shown that
snippets and titles notably affect the user's decision to click
on a specific page \cite{clarke2007influence,joachims2007evaluating,lewandowski2008retrieval,cutrell2007you}.

Unlike previous work, we propose
a metric tailored to the search engine similarity problem that
takes into account diverse criteria as to the rankings and
the content of web results.
Our combined ranking-content
metric aims to return more expressive, objective and robust
similarity scores. It also aims to distinguish important differences
in search engine behavior that are not apparent with other
rank-distance metrics, while viewing each search result as
it is; a unified piece of information.
Furthermore, to the best of our knowledge,
none of the existing studies on search engine similarity
include privacy-friendly search engines, such as DuckDuckGo.

\section{The Metric}
\label{sec:metric}

We introduce a new metric, which we call $T$, to study search engine
similarity. In Section~\ref{ssec:criteria} we formulate the problem
that the metric aims to resolve and the criteria that it should meet;
in Sections~\ref{ssec:starting-point}--\ref{ssec:the-metric} we
develop metric $T$ step-by-step. Then, in
Section~\ref{ssec:comparison} we compare it to other existing metrics.

\subsection{Problem Formulation}
\label{ssec:criteria}
In what follows, we assume that for two search engines $A$ and $B$ we
have two lists $R_A = [a_1, a_2, a_3, \ldots , a_n]$ and
$R_B = [b_1, b_2, b_3, \ldots, b_n]$ of the ranked top $n$ results of
search engine $A$ and search engine $B$ respectively. We denote the
$i^{th}$ element of $R_A$ with $R_A[i]$,
and similarly for $R_B$.

Typically, responses produced by search engines consist of the
identifier of the result's web location ({\sc url}), a result title,
and a snippet describing the page content.
With the evolution of search engines and the
change in user experience, the sequential order
of the search results is not the only factor that determines user
interaction patterns. Specifically, 
snippets and titles are significantly involved
in the user's decision to click
on a specific page \cite{clarke2007influence,joachims2007evaluating,lewandowski2008retrieval,cutrell2007you}. Therefore, a search
engine comparison should take into account all three aspects in order
to accurately appraise engine similarity.

\paragraph{Motivating Example}
To further highlight the importance of snippets and titles, consider
Table~\ref{table:steven} that shows the top result returned by Google
and Bing for the query ``Steven Wilson''. Although search engines
agree in the ordering (i.e., both results point to
\url{http://stevenwilsonhq.com/sw/}), they produce completely
different snippets. Depending on user's search criteria, one snippet
might be more effective on attracting user clicks than the other one.
For instance, the snippet produced by Bing focuses on the artist's
favourite film directors, so it might be a good snippet when users
search for general information about the artist. On the other hand,
the snippet of Google announces his new album release; thus, it gives
emphasis on music news.

\begin{table}[h]
	\centering
	\caption{The top result retrieved for the query ``Steven
		Wilson''.}
	\resizebox{1\linewidth}{!}{%
		\begin{tabular}{lll}
			\textbf{}
			& \multicolumn{1}{c}{\textbf{Bing}}
			& \multicolumn{1}{c}{\textbf{Google}} \bigstrut \\ \cline{2-3}
			\multicolumn{1}{l|}{\textbf{Position}}
			& \multicolumn{1}{c|}{1}
			& \multicolumn{1}{c|}{1} \bigstrut \\ \cline{2-3} 
			\multicolumn{1}{l|}{\textbf{URL}}
			& \multicolumn{1}{p{5.2cm}|}{\url{http://stevenwilsonhq.com/sw/}}
			& \multicolumn{1}{p{5.2cm}|}{\url{http://stevenwilsonhq.com/sw/}} \bigstrut \\ \cline{2-3} 
			\multicolumn{1}{l|}{\textbf{Snippet}}
			& \multicolumn{1}{p{5cm}|}{
				Steven is a film aficionado, and
				frequently cites cinema as one
				of the key inspirations for his music.
				Some of this favourite directors include
				Stanley Kubrick, David Lynch, Ben Wheatley,
				Jonathan Glazer, Shane Meadows and
				Christopher Nolan.}
			& \multicolumn{1}{p{5cm}|}{
				The official website for
				songwriter/producer Steven Wilson.
				New live album/film `Home Invasion:
				In Concert at the Royal Albert Hall' is out now!} \bigstrut \\ \cline{2-3} 
		\end{tabular}
	}
	\setlength{\belowcaptionskip}{5pt}
	\label{table:steven}
\end{table}

\paragraph{Criteria} As the ranking of search engine results does not
fully capture their similarities, we need a more comprehensive
affinity metric. This should meet the following four criteria:
\begin{enumerate}[label={\bf C\arabic*}]
\item The number of common elements (results). The more elements
  search engine $A$ and $B$ share in their top $n$ results, the more
  similar they are.
\item The distance of common elements. If an item appears in the
  results of both $A$ and $B$, the affinity of $A$ and $B$ decreases
  as the distance of the element in the two result lists increases.
\item The importance of agreement decreases as we go down in the
  results lists. For example, agreement at the top result is more
  important than that at the third or fourth result.
\item If two search engines are similar, they produce similar titles
  and snippets, apart from returning similar results in a similar
  order.
\end{enumerate}

\subsection{Starting Point}
\label{ssec:starting-point}

As a starting point to define a metric for search engine affinity, we
take the Jaro-Winkler distance, a variant of the Jaro
distance~\cite{jaro1989advances}, which was applied mainly to the
record linkage problem, and whose goal is to compute string similarity
based on the common elements and the number of transpositions between
them~\cite{winkler1990string}. The Jaro distance of two strings $S_1$
and $S_2$ is given by:
\begin{equation}\label{eq:2}
d_{j}=\left\{{\begin{array}{ll}0 & {\text{if }}m = 0\\
    \frac{1}{3}\left(\frac{m}{|S_{1}|} + \frac{m}{|S_{2}|} +
    \frac{m-t}{m}\right) & {\text{otherwise}}
              \end{array}}\right.
\end{equation}
In the above, $m$ is the number of matching characters and $t$ denotes
the number of transpositions. Two characters are considered matching
if they are the same and their positions do not differ by more than
$(\max(|S_1|, |S_2|)/2) - 1$. The number of transpositions is defined
as half the number of matching characters that are in different order
in the two strings.

The Jaro-Winkler distance extends the Jaro distance by boosting it
using a scaling factor $p$ when the first $l$ characters match
exactly:
\begin{equation}\label{eq:1}
    d_w = d_j + (l \times p \times (1-d_j))
\end{equation}

In order to take into account the snippets and titles returned by the
search engines, we adjust the Jaro-Winkler distance as follows:
\begin{equation}\label{eq:mes}
S=\left\{{\begin{array}{ll}0 & {\text{if }}m = 0\\
    \frac{1}{3n + 1}\left(3m + 1 - a \cdot s -
    b \cdot h - c \cdot t\right)
    & {\text{otherwise}} \end{array}}\right.
\end{equation} where $n$ denotes the common length of the two result sets,
$m = |R_A \cap R_B|$ is the number of common elements,
$t$ is the penalty from transpositions,
$s$ is the penalty from the differences between snippets,
$h$ is the penalty from the differences between titles,
and $a$, $b$, $c$ $\in$ $[0,1]$
are weights attached to the penalties accrued from
snippets, titles, and transpositions respectively.

Note that we compute the ratio of penalties $m-a \cdot s$ and
$m-b \cdot h$ to the length $n$ of results lists rather than the
number of matching elements $m$, which is proposed by Jaro's metric.
This gives us a more reliable estimation of the affinity between
lists. For example, suppose we compare a pair of result rankings of
length $n=10$ and we get the number of matching elements as $m=2$.
According to equation~\ref{eq:2}, if $t=0$ then the term
$\frac{m-t}{m}$ is equal to~1 and it contributes $\frac{1}{3}$ to the
overall similarity, which is a high number, considering the low number
of matching items (two).
Also, we use $m / n$ instead of $m/|S_1| + m/|S_2|$, as $R_A$ and
$R_B$ have a common length $n$.

\subsection{Calculation of Penalties}

\paragraph{Transpositions} To compute transpositions, we take the
sum of the absolute differences of the positions of elements appearing
in both lists. This is a variation of the deviation distance described
by Ronald~\cite{ronald1998}. For lists $R_A$ and $R_B$, the penalty is
computed as follows, where $\sigma(R, a)$ is the position of $a$ in
list $R$:
\[
    t = \frac{
      \sum\limits_{c \in R_A \cap R_B}{|\sigma(R_A, c) - \sigma(R_B, c)|}}
      {t_{\max}}
\]
This penalty is normalized on its upper bound.
It can be proven that in the case of two
lists of length $n$ the upper bound for transpositions
of $|R_A \cap R_B|$ is:
\[
    t_{\max} = \sum_{i=1}^{|R_A \cap R_B|} \phi(i, n)
\] where
\[
    \phi(i, n)=
    \begin{cases}
        n + 1 - i, & \text{if } i = 2k, k \in \mathbb Z^{*}\\
        n - i, & \text{otherwise}
    \end{cases}
\]

\paragraph{Snippets and Titles}

The process of evaluating the penalties related to snippets and titles
is common for both. We examine the sentences $S_1$, $S_2$ of snippets
and titles that are produced by search engines $A$ and $B$ for a
shared result. Then, we tokenize sentences $S_1, S_2$ and eliminate
all stopwords as well as query terms. We get the union of all
tokenized words that appeared in the two sentences and we calculate
the corresponding frequencies, forming two vectors $V_1, V_2$, where
these two vectors represent the actual snippets or titles. We then
compute the cosine distance of the two vectors
$d_s = 1 - \cos(V_1, V_2)$. The overall penalty is computed by
iterating and repeating this process for all common results and then
summing all distances.

\subsection{Similarity Boosting}

The Jaro-Winkler metric treats all explicit matches at the first $l$
characters of strings equally (recall equation~\ref{eq:1}). We,
however, require a descending significance for agreement as we go down
the list of results. To do that, we increase the value of $S$
(equation~\ref{eq:mes}) using weights $w_i$ when there are common
results in positions $1 \le i \le r \le n$. This follows our third
criterion, that exact or adjacent matches are more important at the
beginning of results lists rather than the end. This adjustment
differs from the Jaro-Winkler metric in two ways.

First, the increase is not determined solely by the length of the
matching prefix. For instance, imagine that we compare two vectors,
$V_1= [a, b, c, d, e, f, g], V_2 = [h, b, c, d, e,f, g]$. The
Jaro-Winkler distance does not raise the precomputed Jaro value $d_j$
because there is no matching prefix between $V_1$ and $V_2$. Our
metric rewards agreement at positions up to $r$, e.g., $r=3$ in our
example, therefore, the matches at the second and third items of
$V_1$, $V_2$ increase the computed value.

Second, we also consider that the importance of positions has a
descending order so that each position has an individual contribution
to the total measure. In particular, we define a totally ordered set
$(W, >), |W| = r$, so for each $w_i \in W$:
\begin{equation}\label{eq:weights}
    w_1 > w_2 > w_3 > \ldots > w_r
\end{equation} where $w_1$ is applied in case
of agreement at the first position, $w_2$ at the
second, and so forth.

\subsection{The Metric $T$}
\label{ssec:the-metric}

The final metric of similarity $T$ combines the number of overlapping
results as well as ordering, snippets, and titles of results, and it
is given by:
\begin{equation}\label{eq:boost}
    T = S + \sum_{i=1}^{r} x_iw_i (1 - S)
\end{equation} where
\[
    x_i=
    \begin{cases}
        0, & \text{if } R_{A}[i] \neq R_{B}[i]\\
        1, & \text{otherwise}
    \end{cases}
\]

$T$ meets all four criteria of Section~\ref{ssec:criteria}. The
calculation of overlapping items, $m$, fulfils C1\@. The computation
of the penalty $t
$ fulfils C2, whereas boosting satisfies criterion C3\@.
Finally, $a\cdot s$ and $b \cdot h$ cover C4.

\subsection{Comparison with Other Metrics}
\label{ssec:comparison}
\begin{table*}[th]
\centering
\caption{Results for the comparisons
         described in Section~\ref{ssec:comparison}.}
\renewcommand{\arraystretch}{1.2}
\label{table1}
\resizebox{0.8\linewidth}{!}{%
\begin{tabular}{lccccccc}
  & \makecell{\tt\small a\,b\,c\,d\,e\,f\\\tt a\,g\,h\,i\,j\,k} 
  & \makecell{\tt\small a\,b\,c\,d\,e\,f\\\tt a\,b\,c\,g\,h\,i}
  & \makecell{\tt\small a\,b\,c\,d\,e\,f\\\tt g\,h\,i\,d\,e\,f}
  & \makecell{\tt\small a\,b\,c\,d\,e\,f\\\tt d\,e\,f\,a\,b\,c}
  & \makecell{\tt\small a\,b\,c\,d\,e\,f\\\tt a\,b\,c\,d\,f\,e}
  & \makecell{\tt\small a\,b\,c\,d\,e\,f\\\tt a\,b\,c\,f\,e\,d}
  \\ \hline
\multicolumn{1}{|l|}{Spearman's footrule} & 1.0 & 1.0 & 1.0 & 0.0 & 0.89 & \multicolumn{1}{c|}{0.78}\\ \hline
\multicolumn{1}{|l|}{Kendall's tau} & 1.0 & 1.0 & 1.0 & 0.85 & 0.98 & \multicolumn{1}{c|}{0.95}\\ \hline
\multicolumn{1}{|l|}{$G$} & 0.29 & 0.71 & 0.57 & 0.29 & 0.95 & \multicolumn{1}{c|}{0.90}\\ \hline
\multicolumn{1}{|l|}{$M$} & 0.48 & 0.82 & 0.36 & 0.18 & 0.98 & \multicolumn{1}{c|}{0.97}\\ \hline
\multicolumn{1}{|l|}{Jaro-Winkler} & 0.44 & 0.77 & 0.67 & 0.0 & 0.96 & \multicolumn{1}{c|}{0.96}\\ \hline
\multicolumn{1}{|l|}{$T$} & $[0.24, 0.33]$ &
$[0.46, 0.68]$ & [$0.25, 0.55]$ &
$[0.32, 0.48]$ & $[0.59, 0.89]$ &
\multicolumn{1}{c|}{$[0.57, 0.88$]}\\ \hline
\end{tabular}%
}
\end{table*}

Many metrics have been proposed to evaluate the similarity
of search engine results. However, most of them
focus solely on the ranking of the search results.
The concept behind our proposed metric is to
estimate search engine similarity by incorporating the
semantic features of the search results
into a rank-distance approach.
The dual nature of our metric enables
more expressive and robust
similarity scores and distinguishes important
differences in search engine behavior that are
not apparent with other rank-distance metrics.
In addition, its content-awareness aims to better
reflect the actual user experience, as it has evolved
along with the developments in search engines.

To examine
the behavior of our metric, we use a synthetic example to contrast it
with other metrics. Specifically, we compare it with Spearman's
footrule and Kendall's tau (modified so that they measure similarity
instead of distance)~\cite{fagin2003comparing}, generalizations of
these metrics~\cite{kumar2010generalized}, the Jaro-Winkler metric,
and the metrics $G$ and $M$ proposed by Bar-Ilan et
al.~\cite{bar2006methods,bar2004dynamics}.

Let $L_1 = [{\tt a\,b\,c\,d\,e\,f}]$ be a list that contains responses
provided by one search engine. We compare $L_1$ with six other results
lists, $L_2=[{\tt a\,g\,h\,i\,j\,k}]$, $L_3=[{\tt a\,b\,c\,g\,h\,i}]$,
$L_4=[{\tt g\,h\,i\,d\,e\,f}]$, $L_5=[{\tt d\,e\,f\,a\,b\,c}]$,
$L_6=[{\tt a\,b\,c\,d\,f\,e}]$, $L_7=[{\tt a\,b\,c\,f\,e\,d]}$.
Table~\ref{table1} presents the output of the comparison between the
lists using different metrics.

For the Jaro-Winkler metric we set $p=0.1$, $l\leq 3$. In metric $T$
we set $a = b = c =1.0$ to penalize differences stemming from
snippets, titles, and transpositions respectively, while we set $r=3$,
$w_1 = 0.15$, $w_2=0.1$, $w_3=0.05$ to reward matches at the first $r$
elements.

Only metric $T$ meets criterion C4 regarding snippets and titles.
Thus, we present a lower and upper bound of our metric for every
comparison. The lower bound corresponds to completely different
snippets and titles among common results. The upper bound corresponds
to snippets and titles that are identical.

Taking the different metrics in turn, in Table~\ref{table1} we see
that Spearman's footrule and Kendall's tau ignore mismatching
elements, and compute similarity using only the common ones along with
their distance, therefore, they do not meet criteria C1 and C3\@.

The Jaro-Winkler metric treats equally the transpositions of
$d \leftrightarrow f$ and $e \leftrightarrow f$ in the comparisons
$(L_1, L_6)$ and $(L_1, L_7)$, even though the former introduces a
greater misplacement of elements. Thus, it violates criterion C2.
Moreover, according to equation~\ref{eq:1}, it does not assign
descending significance to agreements at the prefix of lists, which is
required by criterion C3.

Both the $G$ and $M$ metrics (the $M$ metric to a greater extent)
estimate the similarity of lists by giving more emphasis to the
ranking of items rather than the number of overlapping results. For
example, we notice that even though $L_1$ and $L_5$ share all
elements, the values of $M$ and $G$ show a decreasing importance to
greater ranks, especially at the tail of lists. Also, a match in the
first position, as in comparison $(L_1, L_2)$, contributes $0.48$ to
the overall similarity according to the $M$ metric, which is a great
proportion relative to the number of matching items,
i.e., \emph{only }one out of total six.
In essence, although $M$ and $G$ seem to satisfy criteria
C1--C3, they actually ignore matches or adjacent matches at the end of
lists; in fact, the metric $T$ can subsume $G$ and $M$ by using only the
first $q$ elements, for $q < n$.

Kumar and Vassilvitskii have proposed generalized versions of
Spearman's footrule and Kendall's tau
distances~\cite{kumar2010generalized}; their versions take into
account element weights, position weights, and element similarities in
their calculations. It can be shown that the generalizations
overlook elements that appear only in a
single list and thus miss criterion~C1.

%

\section{Evaluation}

\label{sec:engine-comparison}
We compare three search engines, Google, Bing, and DuckDuckGo
(hereafter DDG), for numerous categories of queries, using our
metric~$T$\@. Google and Bing are the two dominant search engines, and
have been the subject of comparative research. DDG adopts a different
philosophy, placing a premium on user privacy. Although far less
popular than the two market leaders, it has a following among
privacy-conscious users. In our empirical evaluation, we try to answer
the following five research questions:\footnote{All data, results,
	and source code used on our experiments are available
	through \url{https://doi.org/10.5281/zenodo.3980817}.}

\begin{enumerate}[label={\bf RQ\arabic*}, leftmargin=2.1\parindent]
\item Do search engines produce similar web results?
(Section~\ref{sec:similarity})

\item Is the similarity between search engines consistent over time?
(Section~\ref{sec:consistency})

\item Which aspect of web results (i.e., rankings or content)
influences the similarity of search engines the most?
(Section~\ref{sec:impact-criteria})

\item Do search engines produce similar results for different
kinds of search services?
(Section~\ref{sec:nature-queries})

\item How do the results produced by the metric $T$ correlate with
the state-of-the-art? (Section~\ref{ssec:other-approaches})

\end{enumerate}
  \vspace{1mm}
\subsection{Dataset}
\label{sec:dataset}
  \vspace{-6mm}
\begin{table}[h]
\centering
\caption{Query categories}
\renewcommand{\arraystretch}{1.2}
\label{table:queries}
\resizebox{0.50\linewidth}{!}{%
\begin{tabular}{|l|l|}
    \hline
    Books \& Authors        & Drinks \& Food   \\ \hline
    Multinational Companies & Music \& Artists \\ \hline
    Politicians             & Regions          \\ \hline
    Software Technologies   & Sports           \\ \hline
    TV \& Cinema            & Universities     \\ \hline
\end{tabular}%
}
\end{table}

Our dataset consists of around~\nnum{27600} top-10 lists in total.
In order to assemble these search results, we constructed
10 categories of queries (Table~\ref{table:queries}).
Every category contains around 30 queries; from these, 20 where taken
from the U.S.\ version of Google
Trends\footnote{\scriptsize\url{https://www.google.com/trends/topcharts}}
in May 2016 and the rest were selected by us.
Given that we cannot test all possible queries,
we selected queries that real people are likely to use
and affect a large number of users.
Furthermore, to achieve representativeness, we enriched each
category with ten more queries selected by us, in order to
include less popular but not rare queries that reflect
the average search use.

For collecting web results,
we are employing the Bing Web Search
API\footnote{\scriptsize\url{https://azure.microsoft.com/en-us/services/cognitive-services/bing-web-search-api/}},
the Google Custom Search
API\footnote{\scriptsize\url{https://developers.google.com/custom-search/}},
and a web scraper that we developed for DDG. Our approach ensures that
the search engines do not take user history into account, which would
affect the final results~\cite{hannak2013measuring}.
We performed the
queries daily, at the same time, for a period of one month (31 days),
from July to August 2016 and a period of 2 months (61 days) from May
to July 2019. We use both datasets to answer RQ2; for the rest of the
research questions, the analysis of them gave consistent results, so,
for brevity, we will focus on the 2019 dataset here. For each query,
we collected the top~10 results using the American domain of each
engine.

\paragraph{URL Normalization}
Each result contains a {\sc url} that specifies its web location. Two
identical {\sc url}s refer to the same result but the same result
could be pointed to by two different {\sc
  url}s~\cite{schonfeld2006not}. This complicates the analysis of
query responses. Some existing studies address this problem by simple
string comparison of {\sc url}s~\cite{spink2006_2overlap,spink2008overlap}. In our work, though, we applied standard
normalization techniques~\cite{lee2005url}. Moreover, to further
reduce the impact of the {\sc dust} problem~\cite{schonfeld2006not}
(i.e., Different {\sc url}s with Similar Text),
we \emph{resolved} redirect {\sc http} responses (3xx status code)
until we arrived at the final target {\sc url}, which we then used for
our analysis. For example, through this method, we can identify that
both \url{https://en.wikipedia.org/wiki/Perth},
and \url{https://en.wikipedia.org/wiki/Perth,_Western_Australia}
return identical content.

\subsection{RQ1: Similarity of Search Engines}
\label{sec:similarity}

We estimate the similarity between Google, Bing and DDG by employing
metric $T$. Specifically, we compare the web results of every search
engine pair, for each query and each date, resulting in three
two-dimensional similarity arrays $D$; one for each search engine
pair. Each element $D_{ij}$ represents the similarity between the two
search engines in the day $i$ for the query $j$.

\begin{figure}[th]
\centering
\includegraphics[scale=0.40]{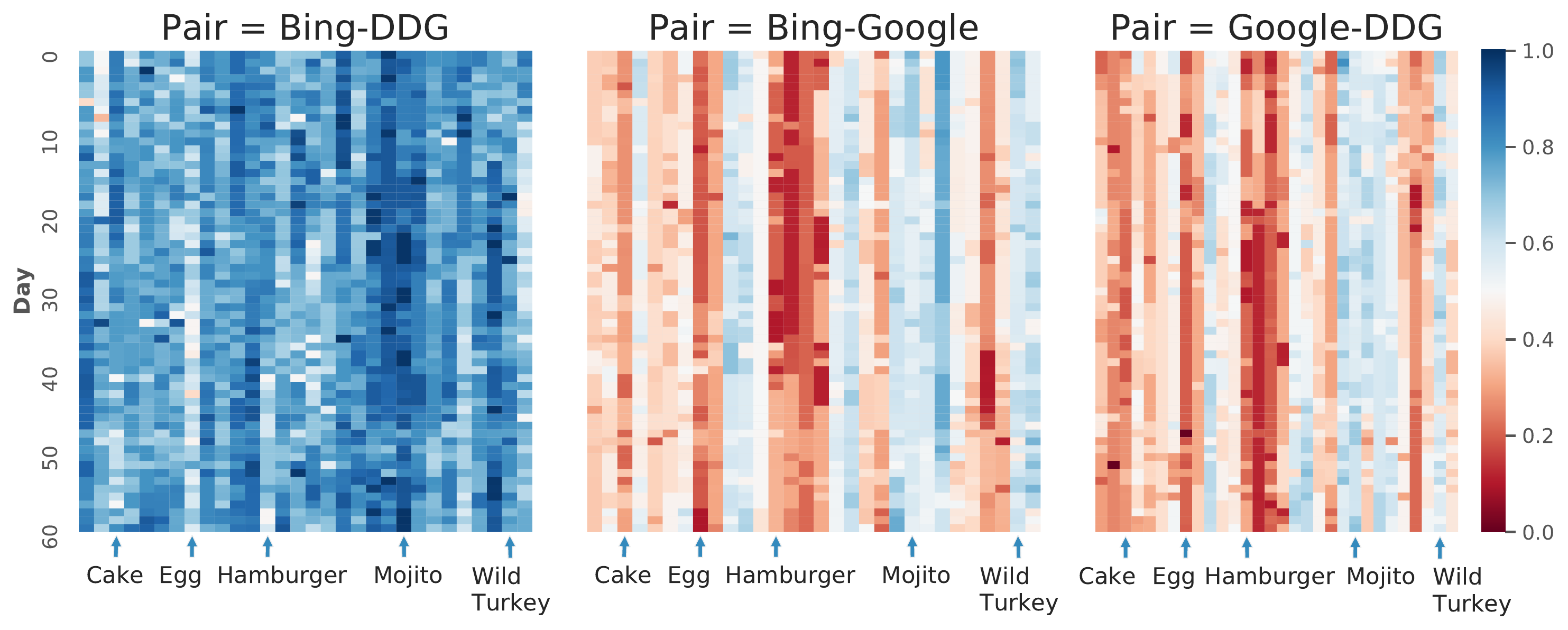}

  \caption{Heatmaps of the
    DDG-Bing, Bing-Google and Google-DDG comparisons.
    There is remarkable
    similarity between DDG and Bing.
    By contrast, Google differs from both
    DDG and Bing to a great extent, for most of the queries.
   } 
  \label{fig:heatmaps}
\end{figure}

Recall from equation~\ref{eq:boost} that we need to define $r$ and the
weights $w_1, w_2, \ldots, w_r$ in order to reward matches at the
first $r$ elements of the ranking lists. In our experiments, we set
$r=5$ and
$W=\{0.15,\allowbreak 0.1,\allowbreak 0.07,\allowbreak
0.03,\allowbreak 0.01\}$; we observed similar tendencies for different
weight assignments. Regarding the importance of result
factors, i.e., snippets, titles, and transpositions,
we set $a=0.8$, $b=1$, $c=0.8$. We use $b = 1$ as the weight for
title penalties, because differences
in titles are rare and in this way we could boost this factor (see
Section~\ref{sec:impact-criteria}).

Figure~\ref{fig:heatmaps} presents the heatmaps of the similarity
arrays $D$ for the queries of the ``Drinks \& Food'' category. These
heatmaps are representative of all the other categories.
Blue cells indicate
cases where search engines are close to each other
(the value of metric $T$ is high),
while red cells reveal dissimilar web results
($T$ is low).

It is easy to see that Bing and DDG give very similar results for the
vast majority of the queries. For example, they returned almost
identical results for the query ``Mojito'', for all the days of our
experiment.
Beyond that, the heatmaps
allow us to pinpoint the cases that do not conform to the general
trend. For example, for the queries
``Egg'' and ``Hamburger'' among DDG and Bing,
the similarity score slightly drops
(observe the corresponding white and orange cells). Overall,
our findings indicate that despite its tiny market share, DDG still
manages to offer a product comparable to that of the market leaders.
The similarity between Bing and DDG could be explained by
the fact that DDG---among other things---employs Bing to get
its results~\cite{DDGsources}.

Moving to Google-Bing and Google-DDG, the results of metric $T$
indicate clear differences as a large number of cells tend to be red.
Specifically, in some cases, the results present
significant contrast, such as the query ``Egg''.
However, there is still a number of queries where the search engines
seem to have a slightly higher degree of resemblance,
e.g.,``Wild Turkey'' and ``Mojito''. Since, not much is known on
how search engine ranking algorithms work, another factor that
affects search engines' similarity is their
index size; in particular, Google has
demonstrated a higher average index size
than Bing~\cite{vandenBosch2016index}.

\noindent
\begin{tcolorbox}
{\bf Finding \#1.}
Google stands apart from Bing and DDG for the
majority of the queries, while the latter two are mostly
indistinguishable from each other.
\end{tcolorbox}

\subsection{RQ2: Consistency of Search Engines}
  \vspace{-3mm}
\label{sec:consistency}

\begin{figure}[ht]
    \centering
    \includegraphics[scale=0.40]{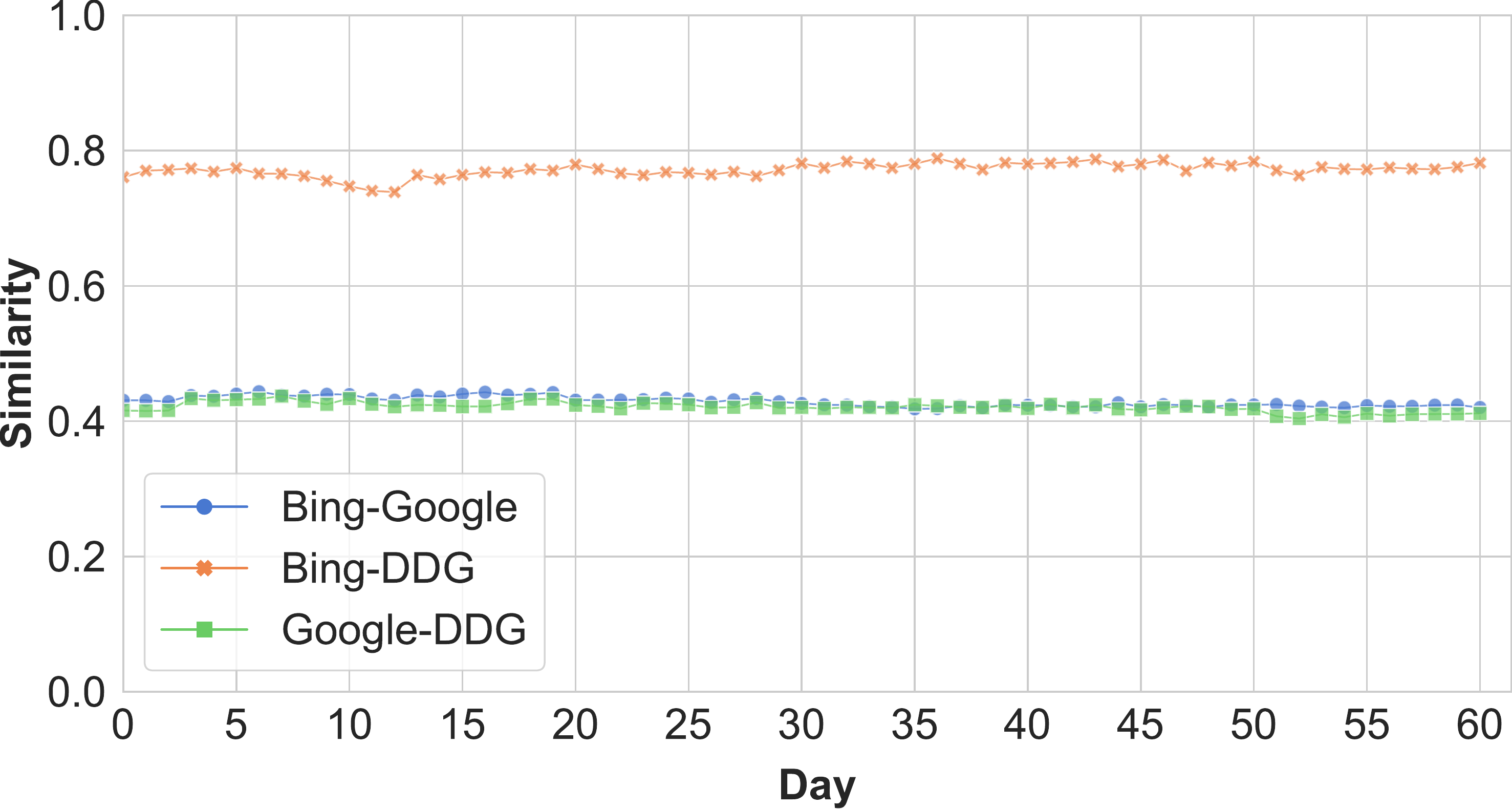}
    \caption{Similarity evolution for a 61-day period in 2019.
    The similarity between all comparison pairs is consistent in
    the short-term with only a small number of trivial fluctuations.
    The similarity between DDG and Bing is almost double than that
    of Bing-Google and Google-DDG.}
    \label{fig:consistency_plot}
\end{figure}
\begin{figure}[ht]
\centering
\includegraphics[scale=0.30]{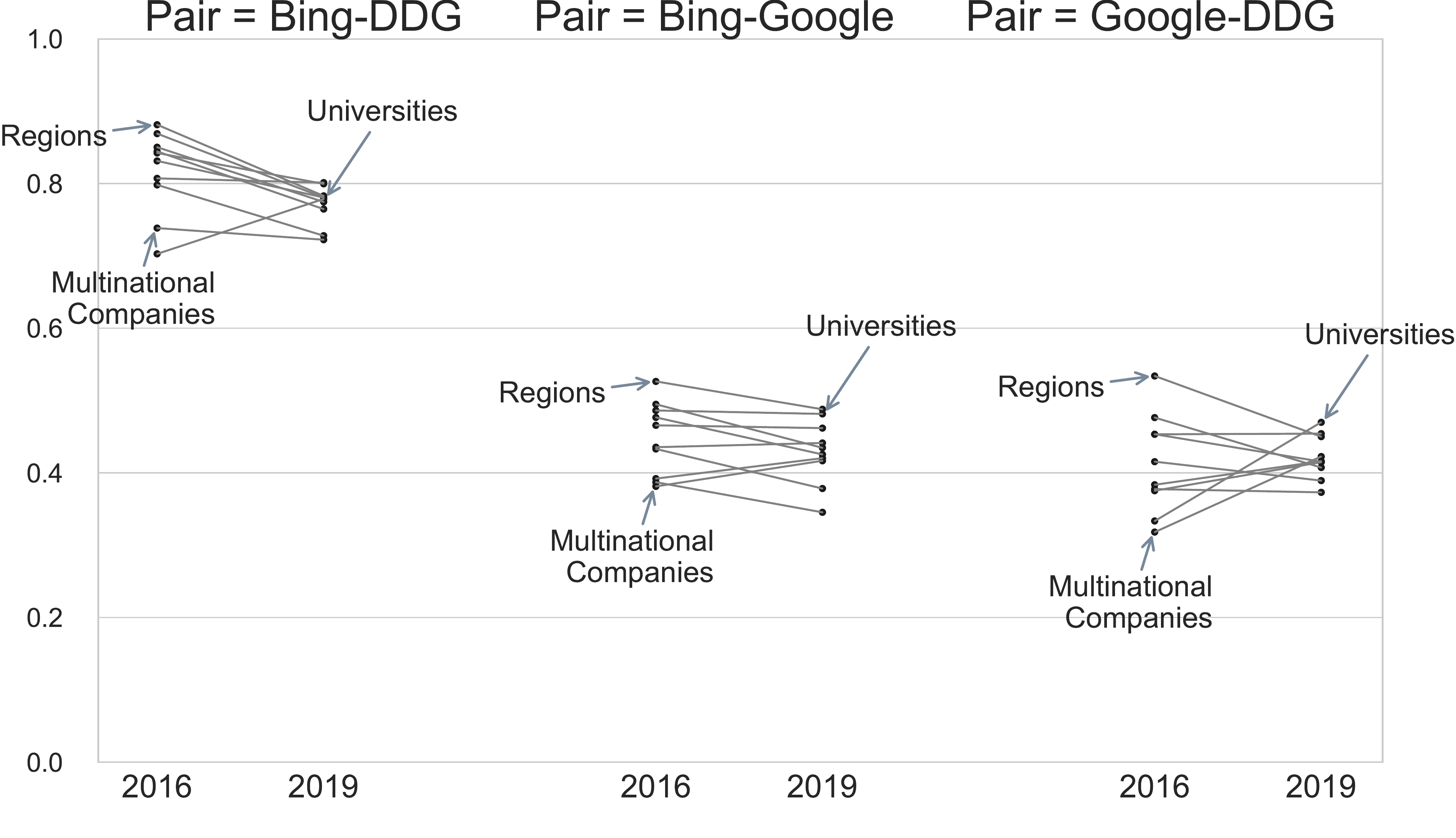}
\caption{Comparison of search engines' similarity in 2016 and 2019.
For every search engine pair, we assess their degree of resemblance
for every query category in 2016 and 2019.
We observe that search engines' similarity does not change
considerably  in the long-term. Bing-DDG is the most
similar pair both in 2016 and 2019.}
\label{fig:year_plot}
\end{figure}

To estimate the consistency of search engine behavior over time, we
calculate the average similarity score, as computed by metric $T$, of
the search engine results for every day and comparison pair. This is
the mean per row of every array $D$. Figure~\ref{fig:consistency_plot}
presents the average similarity of every search engine pair over time.
This figure clearly shows that the affinity of the search engines is
almost constant over time, with only a small number of trivial
fluctuations. The findings from this experiment imply that either the
search engines do not significantly change their behavior or that
their behavior changes in the same way. 

In addition, the plots reveal that the similarity between Bing-DDG is
almost double than that of Bing-Google and Google-DDG, strengthening
our first finding. Also, Bing and DDG seem to behave with a similar
manner when they are compared with Google, because the
corresponding plots (observe the blue and green plots) almost lie on
each other.

\noindent
\begin{tcolorbox}
{\bf Finding \#2.}
The behavior of all the search engines
remains consistent in the short-term.
\end{tcolorbox}
  \vspace{-3mm}

We also examined how the search engines' similarity changes in the
long-term. To do so, we compared the $T$ values between the 2016 and
2019 datasets. Figure~\ref{fig:year_plot} shows the differences in the
similarity of every search engine for each query category from 2016 to
2019. Overall, we see that Bing and DDG moved from being very similar
to slightly less so (their similarity decreases by 7.4\%, on average).
The affinity between of Bing-Google is almost stable (it drops by
\emph{only} 1.6\%, on average). Finally, DDG has come
somewhat closer to
Google, i.e., there is an increase in their similarity by 4.5\% on
average. We inspected the categories, focusing especially
on those that are bucking the trend (e.g.,
``Universities'' in Bing-DDG). The results show that
the increase in the similarity between Google and DDG is
due to changes in DDG's results within this period of time.

Apart from the similarity of search engine pairs between 2016 and
2019, we examined how a search engine changes itself between these two
points in time. Certainly, in three years the world changes and
information sources change as well. Therefore, we do not expect a
search engine to return the same results in 2016 and in 2019 for the
same query. However, although we expect all three search engines to
change, we do not how to what degree they will change. We found that
the average similarity between 2016 and 2019 is 0.37 for DDG, 0.43 for
Bing, and 0.48 for Google; that is, Google changed the least and DDG
the most. DDG's rankings and search algorithms may have been updated
to a greater extent than the other search engines. This may have
supported DDG's relative growth (its market share moved upwards by
274.28\%) over the past three years~\cite{nyt,stat}.

\noindent
\begin{tcolorbox}
{\bf Finding \#3.}
Bing and DDG remain more similar to each other
than Bing-Google and Google-DDG.
Although search engines change individually,
their pairwise similarity is almost stable in the long-term.
\end{tcolorbox}

\subsection{RQ3: Impact of Snippets, Titles, and Transpositions}
\label{sec:impact-criteria}

\begin{figure}
\centering
\includegraphics[scale=0.32]{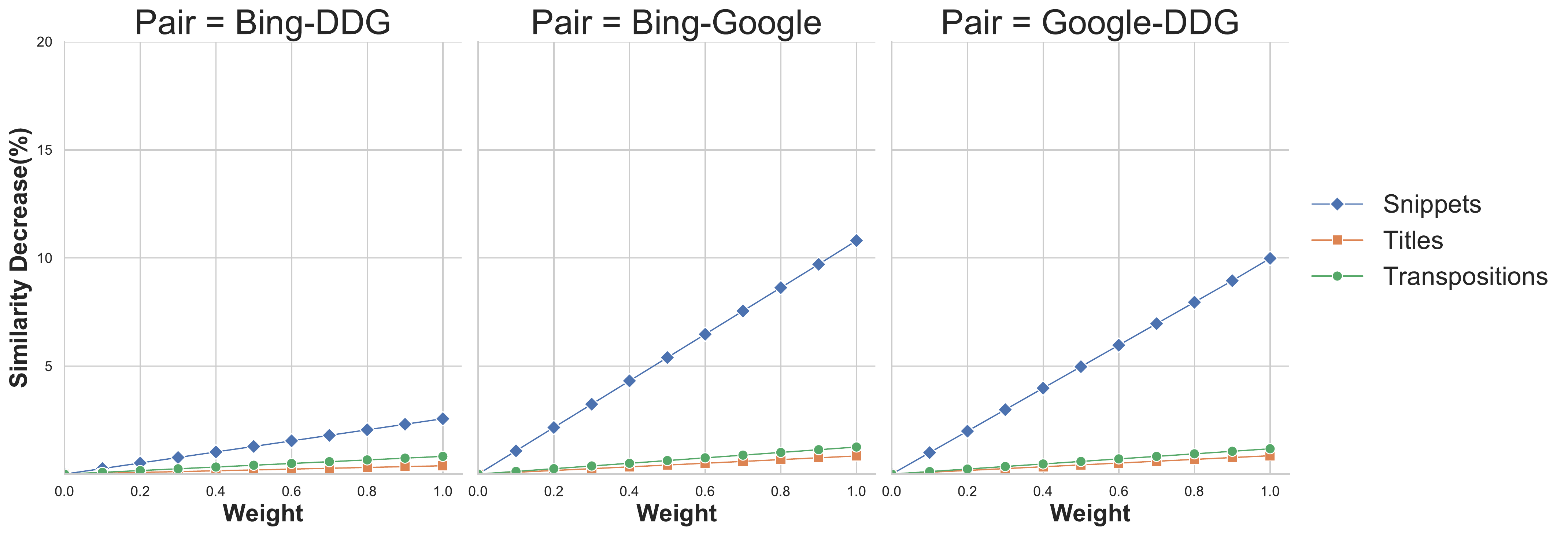}
\caption{The impact of snippets, titles and transpositions
on the similarity of each pair. Every plot shows the percentage decrease in the
similarity of {\it T} for different weight assignments of each factor.
Google seems to construct different snippets compared to DDG
and Bing.}
\label{fig:impact}
\end{figure}

Unlike existing approaches, metric $T$ captures both the ordering
(i.e., transpositions) and the content (i.e., snippets, titles) of
results. Therefore, we can estimate how much each factor contributes
to the differences between search engines. To do so, we instantiate the
metric $T$ with different weights for each factor (recall $a, b, c$
from Equation~\ref{eq:mes}). We first consider the metric
$T_{\mathit{base}}$ as the baseline metric with weights $a=0$, $b=0$,
$c=0$. We compute the average similarity of every comparison pair for
all the queries and days. Conceptually, $T_{\mathit{base}}$ considers
only the number of overlapping results and the agreements at the first
$r=5$ results. Then, we examine the effect of snippets by varying
$a = 0.1$, $0.2$, $\ldots$, $1$ while keeping $b = c = 0$. Similarly,
we examine the effect of titles and transpositions by varying $b$ and
$c$ while keeping the other two weights pegged to zero.

In figure~\ref{fig:impact}, each diagram shows the impact of every
factor on the decrease of $T_{\mathit{base}}$ for each search engine
pair. It is clear that snippets have the biggest impact, by a wide
margin. The difference in transpositions is much smaller, and the
difference in titles minimal, throughout all search engine
comparisons. Google seems to construct different snippets
compared to Bing and DDG, an observation that is consistent
with our motivating example in Section~\ref{ssec:criteria}.
\noindent
\begin{tcolorbox}
{\bf Finding \#4.}
Snippets have the greatest impact on
the differences among all the comparison pairs,
with Google yielding the most distinct ones.
All the search engines tend to place
their common results in adjacent positions.
Finally,
all the search engines produce almost identical titles.
\end{tcolorbox}

\subsection{RQ4: Search Engine Similarity in Different Search Services}
\label{sec:nature-queries}

Apart from standard web search, search engines provide their users
with a list of different services, such as, news, image, and video
search. We investigated whether our findings regarding the similarity
between search engines apply to the news search tab for 2019---we
excluded other services as our metric works on textual results. We
created a set of 30 news queries; 20~of them were taken from the
Google News trends of May 2019 and the remaining~10 were generic news
topics, e.g., ``flood''.

The results show a very low average similarity of 0.12, in contrast
with the average 0.54 similarity of the results from the regular
search. Bing-Google is the pair that exhibits the highest similarity
(0.15). Overall, this dissimilarity can be explained by the ephemeral
nature of the news that requires quick evaluation and leads to daily
ups and downs of topics and content. Also, the ranking algorithm of
the news search results is different than that of the regular search,
certainly for Google~\cite{economist_2019}. Each ranking algorithm
takes into account multiple different parameters, such as human
evaluators ~\cite{economist_2019,filloux_2013}, that can lead to
completely divergent results. As indicated in the work of Agrawal
et al.~\cite{agrawal}, the low similarity between search engines'
results is preferable, especially for informative search, because
users get exposed to diverse views and perspectives (e.g., different
political views).

\noindent
\begin{tcolorbox}
{\bf Finding \#5.}
There is a considerable difference in the results produced by
different search engines' services.
\end{tcolorbox}
\subsection{RQ5: Comparison with Other Approaches}
\label{ssec:other-approaches}
\begin{figure}[ht]
\centering

\includegraphics[scale=0.40]{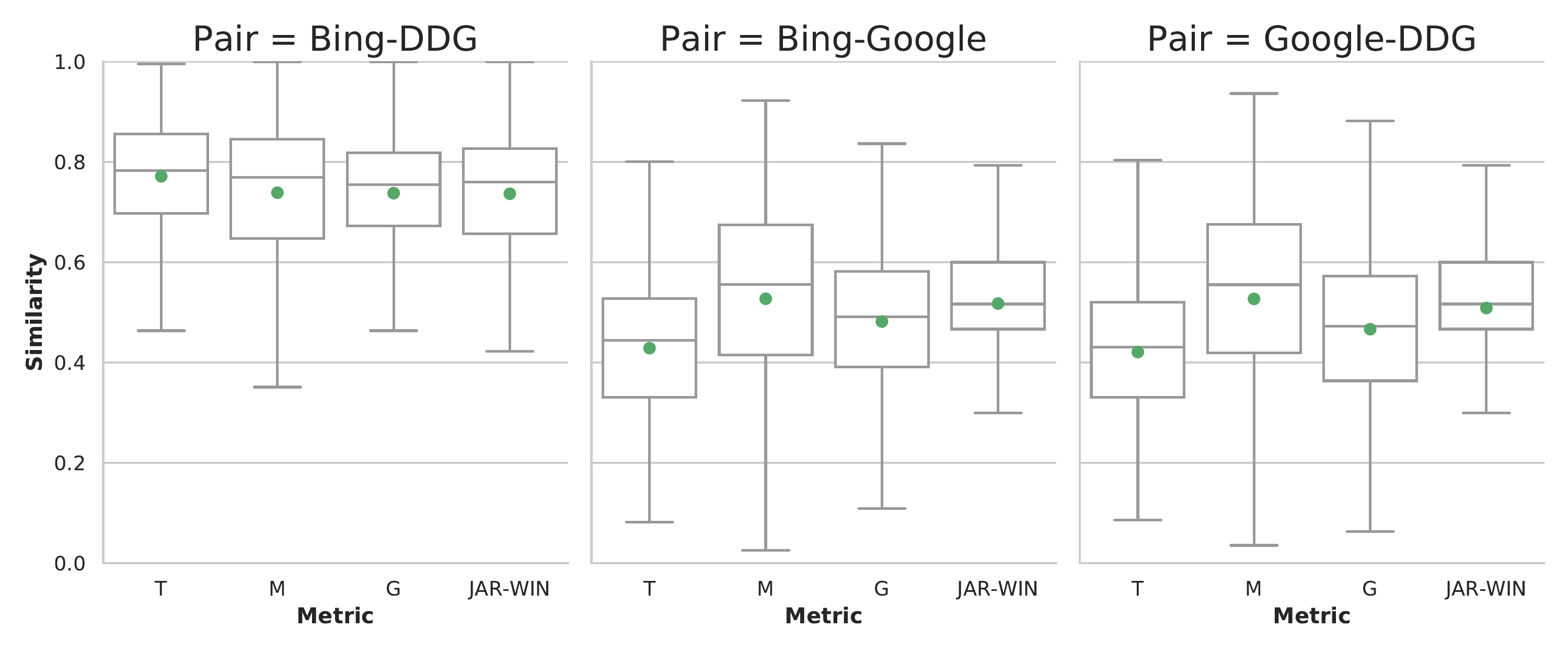}
\caption{Similarity of all search engine pairs
using different metrics. The box plots of metric {\it T}
are lower than those of the other metrics for the Bing-Google
and Google-DDG comparisons, because it effectively takes into account
their significant disagreements as to the content of results.}
\label{fig:metric-boxplot}

\end{figure}
\paragraph{Rankings-Based Approaches}
We study how metric $T$
correlates with three metrics that have been used in search
engine comparisons (Section~\ref{ssec:comparison}). Specifically,
we use the metrics $M$, $G$, and Jaro-Winkler to compute the
similarity between every search engine pair
like we did in Section~\ref{sec:similarity} using metric $T$.

Figure~\ref{fig:metric-boxplot} shows the box plots of the search
engine affinity for each metric. Every box plot contains the median
similarity (horizontal line), the mean similarity (green circle),
along with the maximum and minimum similarity values. The figure
replicates our first finding (Section~\ref{sec:similarity}), that is,
Google seems to produce more unique results when compared to Bing or
DDG, as the corresponding box plots are lower than those of the
Bing-DDG pair. Hence, the results of metric $T$ are consistent with
those of the three aforementioned metrics.

However, the box plots demonstrate that the metric $T$ seems to stand
out from the others, especially in the Bing-Google, Google-DDG pairs.
Specifically, by observing the average and the median similarity, we
see that metric $T$ always produces lower values than the other
metrics. This is explained by the fact that metric $T$
\emph{effectively} captures differences that stem from snippets and
titles in those comparison pairs, which the other metrics ignore
(Section~\ref{sec:impact-criteria}).

\noindent
\begin{tcolorbox}
{\bf Finding \#6.}
Metric $T$, when compared to others, exhibits a consistent behavior.
However, when the content similarity falls, the results of metric $T$
differ from those of the other metrics.
\end{tcolorbox}

\begin{figure}[t]
  \includegraphics[scale=0.40]{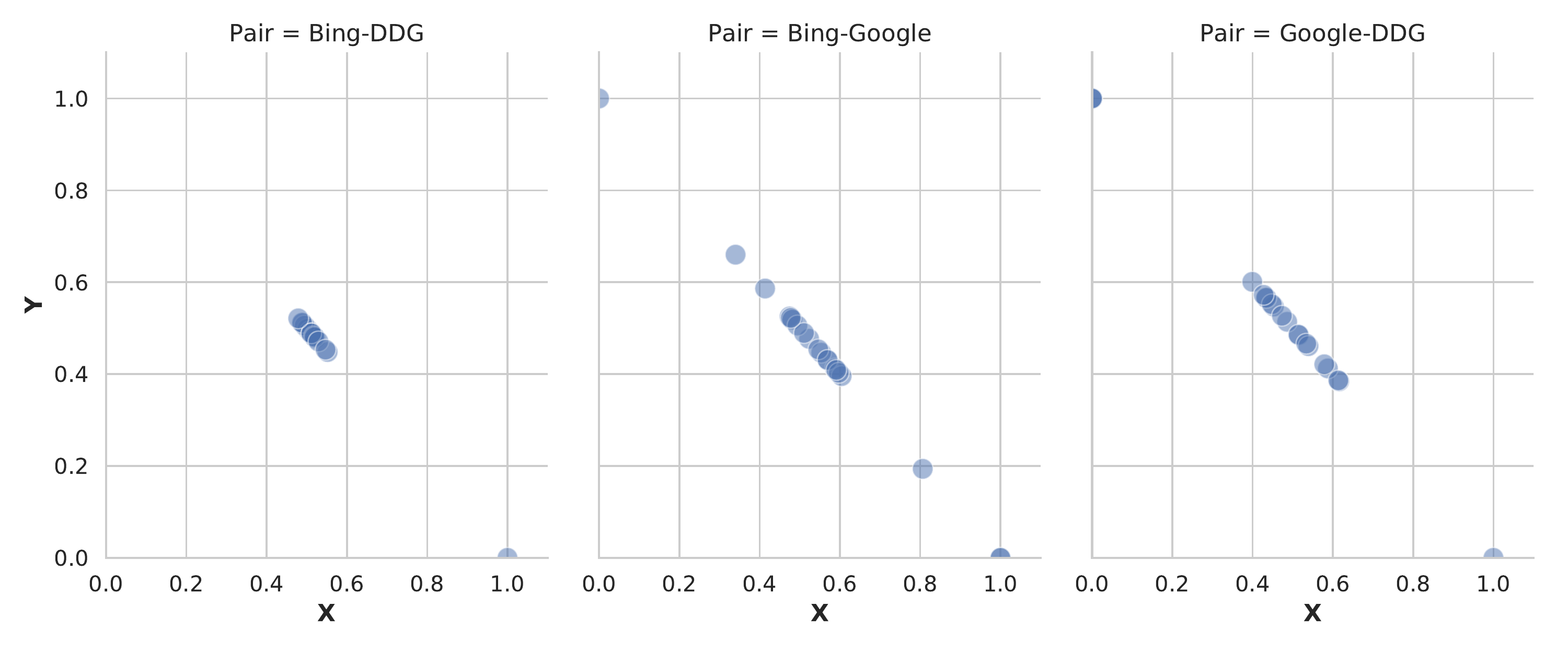}
  \caption{Results of TensorCompare comparing every search engine
    pair. Each point $(x, y)$ represents the contribution of each
    search engine to a specific component. A uniform ratio of
    contribution implies high affinity of the compared engines.
    All the observations (except for one)
    from DDG-Bing lie between $[0.45, 0.55]$,
    whereas there is a number
    of observations located on extreme points (1, 0), (0, 1)
    as shown from the comparisons of Bing-Google and DDG-Google.}
  \label{fig:tc}
\end{figure}

\paragraph{Content-Based Approaches}
Agrawal et al.~\cite{agrawal2015study} have introduced TensorCompare
and CrossLearnCompare, two content-based methods that utilize tensor
decomposition and supervised learning techniques. Both methods exploit
snippets of the search results in order to compare their content, but
without considering their ordering. Nevertheless, we used them in
comparison with our method
to cross-check the metric $T$ with
different content-based methods.

TensorCompare constructs a four mode tensor $X$ (result, query, day,
search engine) and then applies a tensor decomposition algorithm
~\cite{harshman1970foundations} on it to get a sum of factor matrices,
reflecting the corresponding modes of the tensor. CrossLearnCompare
builds a classification model predicting the query that produces a
result of a search engine $A$
given the results of a search engine $B$ as training data. 
The performance of the classifier then indicates how
similar these search engines were. These methods were used to compare
Google and Bing. The study results showed that these search engines
seem to be significantly similar when judged on the content of their
results.


We tried to replicate their findings. In order to test TensorCompare,
we made our own implementation of the Alternating Poisson Regression
(CP APR)~\cite{chi2012tensors} method, which is used for the tensor
decomposition. To ensure the correctness of our implementation we
tested its results against of that of the Matlab TensorToolbox%
\footnote{\scriptsize{
    \url{http://www.sandia.gov/~tgkolda/TensorToolbox/index-1.0.html}}},
using the method described by Albanese et
al.~\cite{albanese2013minerva} in their implementation of the Maximal
Information Coefficient (MIC) metric introduced by Reshef et
al.~\cite{reshef2011detecting}. We followed exactly the steps
described in their study to build the feature space and feed both
methods.

\begin{figure}[t]
    \centering
    \includegraphics[scale=0.45]{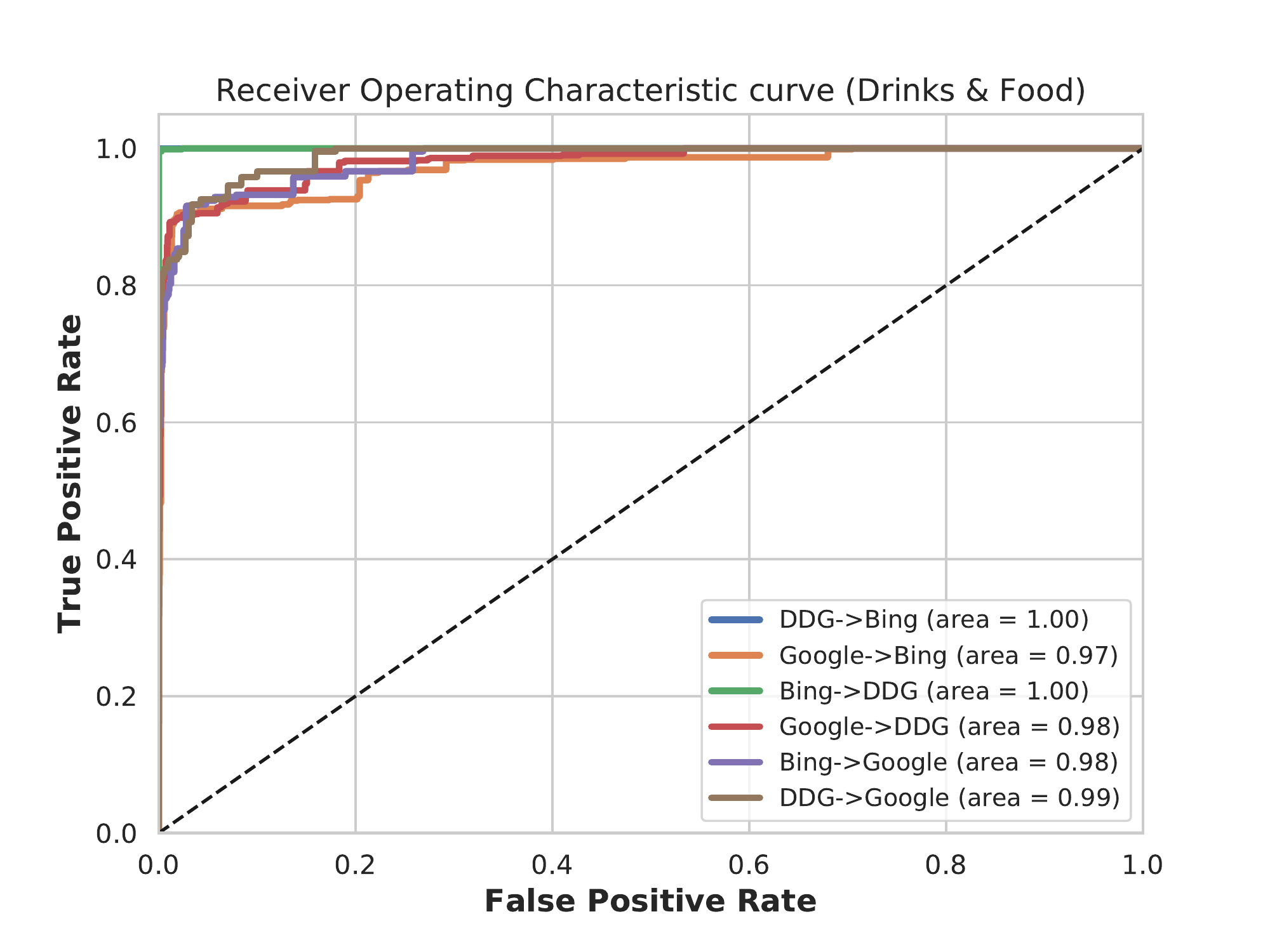}
    \caption{\label{fig:clc} Results of CrossLearnCompare. We trained
      six classifiers, one for every search engine pair, to predict
      queries. The results of the search engines on the right were
      used as training data in order to predict the search engines on
      the left. All Search engines---and in particular Bing and
      DDG---appear almost identical, because of high predictability, as
      shown by Receiver Operating Characteristic (ROC) curves under
      their Area under the Curve (AUC).}
\end{figure}
We ran these methods for every query category, incorporating results
of all the days examined by our experiment.
The TensorCompare results (see
Figure~\ref{fig:tc}) for the ``Drinks \& Food'' category
(we make similar findings for the rest of the categories)
confirm that the Bing-DDG pair is much
more related than the rest.
This underpins our findings about RQ3
where Bing and DDG return snippets
that are very close to each other.

The results of CrossLearnCompare (Figure~\ref{fig:clc}) clash with the
results of the other methods. Specifically, according to the
assessment of Agrawal et al., all search engines seem to behave with
an almost identical manner as shown by the high predictability of
queries, and indeed we replicated that. An explanation for this could
be that CrossLearnCompare actually predicts \emph{queries} and not
\emph{search engines}, which may be distinguishable from each other.
As witnessed in the results obtained by the other methods, predicting
queries rather than search engines means that it ignores the
differences stemming from both the rankings and the snippets (as shown
in Section~\ref{sec:impact-criteria}) of results. In this way,
CrossLearnCompare mistakenly indicates that there is a significant
overlap in the results generated by Bing-Google and
Google-DDG~\cite{agrawal,agrawal2015study}.

However,
notice that---once again---the pair of Bing-DDG
is more predictable than the rest,
meaning that those search engines
generate web results with more similar contents
than Bing-Google and Google-DDG,
as we also find through our metric $T$ (Section~\ref{sec:impact-criteria}).


\subsection{Threats to Validity}

The main threat to external validity is the representativeness
of the selected queries.
To mitigate this threat,
we created a large corpus of
\nnum{27600} lists of top-10 search results. These results
were assembled from
300 unique queries, spanning 10 different topics.
The two-thirds of the queries were
taken from the Google trends of 2016, in order to
study the search engines in use cases that impact
a large number of users.
The rest were selected by us, aiming to include
less popular but not rare queries that better reflect
the average search use. 

Another
threat to external validity could be that we considered only the top
$n$ web results for every query. However, we based our decision on
previous studies of user behavior~\cite{enge2012art,joachims2002optimizing} that have shown that it is more likely for
users to click an item included in the first ten results.

The main threat to internal validity is the design of our metric $T$.
To alleviate this threat, we kept four criteria in mind
(Section~\ref{ssec:criteria}) that are considered very important in
search engine comparisons, and captured both the rankings and the
content of the results. We compared the results of the metric $T$ with
other existing methods (both rankings- and content-based), showing
that the results of our metric are consistent with those found by
others. Another threat comes from the methodology of collecting the
web results. We used the {\sc rest} {\sc api}s of Google and Bing so
that the web results are not biased by our search
history~\cite{hannak2013measuring}. We queried every search engine at
the same time, each day, using the same parameters, employing
standard {\sc url} normalization methods, and resolving {\tt http}
redirections to mitigate the {\sc dust} problem.

\section{Conclusions and Discussion}
\label{sec:conclusions-discussion}

In this work, we introduce a novel similarity metric for
search engine comparison that combines the
rankings of results and their semantic presentation.
In contrast to the existing ranking-based or content-based
approaches, our metric aims to be more expressive, robust and
objective, following the
aggregation of heterogeneous information into search results and
the emergence of new user interaction patterns. Thus,
it effectively captures differences that stem from snippets
and titles, which the other metrics ignore.

By employing our metric, we were
able to track engine similarity on both content and ranking
across time, for a large and broad number of queries.
Our results indicate that Google stands apart from Bing and
DuckDuckGo, but these two are largely indistinguishable.
The performance of DuckDuckGo
may run counter to many expectations, taking into account the
comparatively vast
disparity of its resources. In our study we queried search engines
without taking into account the user history. It is possible that
when user history is employed, Bing would differ measurably
from DuckDuckGo. Still, Google manages to differ from both Bing and
DuckDuckGo even when it does not leverage personalized data.

Lately, search engines have started producing summaries,
overviews, and compelling navigational aids,
calling for more flexible comparison methodologies.
Our approach consists a first
step towards this direction, but the
incorporation of semantically-rich features in search engine similarity
measures seems a promising area for future research. \vspace{-1mm}

\subsubsection{Acknowledgments}
This work was supported by the European Union’s Horizon 2020 research and innovation program ``FASTEN'' under grant agreement No 825328.
\vspace{-1mm}

%

\bibliographystyle{splncs04}
\bibliography{sesim}

\begin{thebibliography}{10}
\providecommand{\url}[1]{\texttt{#1}}
\providecommand{\urlprefix}{URL }
\providecommand{\doi}[1]{https://doi.org/#1}

\bibitem{agrawal2015study}
Agrawal, R., Golshan, B., Papalexakis, E.: A study of distinctiveness in web
  results of two search engines. In: Proceedings of the 24th International
  Conference on World Wide Web. pp. 267--273. ACM (2015)

\bibitem{agrawal}
Agrawal, R., Golshan, B., Papalexakis, E.: Whither social networks for web
  search? In: Proceedings of the 21th ACM SIGKDD International Conference on
  Knowledge Discovery and Data Mining. pp. 1661--1670. KDD '15, ACM, New York,
  NY, USA (2015). \doi{10.1145/2783258.2788571},
  \url{http://doi.acm.org/10.1145/2783258.2788571}

\bibitem{albanese2013minerva}
Albanese, D., Filosi, M., Visintainer, R., Riccadonna, S., Jurman, G.,
  Furlanello, C.: Minerva and minepy: a {C} engine for the {MINE} suite and its
  {R}, {P}ython and {MATLAB} wrappers. Bioinformatics  \textbf{29}(3),
  407--408 (2013)

\bibitem{arguello2011aggregate}
Arguello, J., Diaz, F., Callan, J.: Learning to aggregate vertical results into
  web search results. In: Proceedings of the 20th ACM International Conference
  on Information and Knowledge Management. p. 201–210. CIKM ’11,
  Association for Computing Machinery, New York, NY, USA (2011).
  \doi{10.1145/2063576.2063611}, \url{https://doi.org/10.1145/2063576.2063611}

\bibitem{arguello2009vertical}
Arguello, J., Diaz, F., Callan, J., Crespo, J.F.: Sources of evidence for
  vertical selection. In: Proceedings of the 32nd International ACM SIGIR
  Conference on Research and Development in Information Retrieval. p.
  315–322. SIGIR ’09, Association for Computing Machinery, New York, NY,
  USA (2009). \doi{10.1145/1571941.1571997},
  \url{https://doi.org/10.1145/1571941.1571997}

\bibitem{bailey2010whole}
Bailey, P., Craswell, N., White, R.W., Chen, L., Satyanarayana, A., Tahaghoghi,
  S.: Evaluating whole-page relevance. In: Proceedings of the 33rd
  International ACM SIGIR Conference on Research and Development in Information
  Retrieval. p. 767–768. SIGIR ’10, Association for Computing Machinery,
  New York, NY, USA (2010). \doi{10.1145/1835449.1835606},
  \url{https://doi.org/10.1145/1835449.1835606}

\bibitem{bar1999search}
Bar-Ilan, J.: Search engine results over time: {A} case study on search engine
  stability. Cybermetrics  \textbf{2}(3), ~1 (1999)

\bibitem{bar2007user}
Bar-Ilan, J., Keenoy, K., Yaari, E., Levene, M.: User rankings of search engine
  results. Journal of the American Society for Information Science and
  Technology  \textbf{58}(9),  1254--1266 (2007)

\bibitem{bar2004dynamics}
Bar-Ilan, J., Levene, M., Mat-Hassan, M.: Dynamics of search engine
  rankings-{A} case study. In: Proceedings of the 3rd International Workshop on
  Web Dynamics (2004)

\bibitem{bar2006methods}
Bar-Ilan, J., Mat-Hassan, M., Levene, M.: Methods for comparing rankings of
  search engine results. Computer networks  \textbf{50}(10),  1448--1463 (2006)

\bibitem{krishna1998overlap}
Bharat, K., Broder, A.: A technique for measuring the relative size and overlap
  of public web search engines. Computer Networks and ISDN Systems
  \textbf{30}(1),  379–388 (1998)

\bibitem{bian2010query}
Bian, J., Liu, T.Y., Qin, T., Zha, H.: Ranking with query-dependent loss for
  web search. In: Proceedings of the Third ACM International Conference on Web
  Search and Data Mining. p. 141–150. WSDM ’10, Association for Computing
  Machinery, New York, NY, USA (2010). \doi{10.1145/1718487.1718506},
  \url{https://doi.org/10.1145/1718487.1718506}

\bibitem{vandenBosch2016index}
van~den Bosch, A., Bogers, T., de~Kunder, M.: Estimating search engine index
  size variability: a 9-year longitudinal study. Scientometrics
  \textbf{107}(2),  839--856 (May 2016). \doi{10.1007/s11192-016-1863-z}

\bibitem{cardoso2011google}
Cardoso, B., Magalh{\~a}es, J.: Google, {B}ing and a new perspective on ranking
  similarity. In: Proceedings of the 20th ACM international conference on
  Information and knowledge management. pp. 1933--1936. ACM (2011)

\bibitem{danqi2012blue}
Chen, D., Chen, W., Wang, H., Chen, Z., Yang, Q.: Beyond ten blue links:
  Enabling user click modeling in federated web search. In: Proceedings of the
  Fifth ACM International Conference on Web Search and Data Mining. p.
  463–472. WSDM ’12, Association for Computing Machinery, New York, NY, USA
  (2012). \doi{10.1145/2124295.2124351},
  \url{https://doi.org/10.1145/2124295.2124351}

\bibitem{chi2012tensors}
Chi, E.C., Kolda, T.G.: On tensors, sparsity, and nonnegative factorizations.
  SIAM Journal on Matrix Analysis and Applications  \textbf{33}(4),  1272--1299
  (2012)

\bibitem{rosenthal1999search}
Chu, H., Rosenthal, M.: Search engines for the world wide web: A comparative
  study and evaluation methodology. Proceedings of the ASIS Annual Meeting
  \textbf{33},  127--135 (1996)

\bibitem{clarke2007influence}
Clarke, C.L., Agichtein, E., Dumais, S., White, R.W.: The influence of caption
  features on clickthrough patterns in web search. In: Proceedings of the 30th
  annual international ACM SIGIR conference on Research and development in
  information retrieval. pp. 135--142. ACM (2007)

\bibitem{collier2014information}
Collier, J.H., Konagurthu, A.S.: An information measure for comparing top k
  lists. In: e-Science (e-Science), 2014 IEEE 10th International Conference on.
  vol.~1, pp. 127--134. IEEE (2014)

\bibitem{cutrell2007you}
Cutrell, E., Guan, Z.: What are you looking for?: an eye-tracking study of
  information usage in web search. In: Proceedings of the SIGCHI conference on
  Human factors in computing systems. pp. 407--416. ACM (2007)

\bibitem{marchionini1996search}
Ding, W., Marchionini, G.: A comparative study of web search service
  performance. Proceedings of the ASIS Annual Meeting  \textbf{33},  136--142
  (1996)

\bibitem{DDGsources}
DuckDuckGo: Duckduckgo sources.
  \url{https://help.duckduckgo.com/results/sources/} (2019), online accessed;
  07 August 2019

\bibitem{economist_2019}
Economist, T.: Seek and you shall find: {Google} rewards reputable reporting,
  not left-wing politics  (Jun 2019),
  \url{https://www.economist.com/graphic-detail/2019/06/08/google-rewards-reputable-reporting-not-left-wing-politics}

\bibitem{enge2012art}
Enge, E., Spencer, S., Fishkin, R., Stricchiola, J.: The art of {SEO}. "
  O'Reilly Media, Inc." (2012)

\bibitem{fagin2003comparing}
Fagin, R., Kumar, R., Sivakumar, D.: Comparing top $k$ lists. SIAM Journal on
  Discrete Mathematics  \textbf{17}(1),  134--160 (2003)

\bibitem{filloux_2013}
Filloux, F.: Google news: the secret sauce. The Guardian  (Feb 2013),
  \url{https://www.theguardian.com/technology/2013/feb/25/1}

\bibitem{stat}
GlobalStats, S.: Statcounter globalstats. \url{http://gs.statcounter.com}
  (2019), online accessed; 06 August 2019

\bibitem{gordon1999finding}
Gordon, M., Pathak, P.: Finding information on the {W}orld {W}ide {W}eb: the
  retrieval effectiveness of search engines. Information Processing \&
  Management  \textbf{35}(2),  141--180 (1999)

\bibitem{hannak2013measuring}
Hannak, A., Sapiezynski, P., Molavi~Kakhki, A., Krishnamurthy, B., Lazer, D.,
  Mislove, A., Wilson, C.: Measuring personalization of web search. In:
  Proceedings of the 22nd international conference on World Wide Web. pp.
  527--538. ACM (2013)

\bibitem{harshman1970foundations}
Harshman, R.A.: Foundations of the {PARAFAC} procedure: {M}odels and conditions
  for an" explanatory" multi-modal factor analysis  (1970)

\bibitem{jaro1989advances}
Jaro, M.A.: Advances in record-linkage methodology as applied to matching the
  1985 census of {T}ampa, {F}lorida. Journal of the American Statistical
  Association  \textbf{84}(406),  414--420 (1989)

\bibitem{joachims2002optimizing}
Joachims, T.: Optimizing search engines using clickthrough data. In:
  Proceedings of the eighth ACM SIGKDD international conference on Knowledge
  discovery and data mining. pp. 133--142. ACM (2002)

\bibitem{joachims2007evaluating}
Joachims, T., Granka, L., Pan, B., Hembrooke, H., Radlinski, F., Gay, G.:
  Evaluating the accuracy of implicit feedback from clicks and query
  reformulations in web search. ACM Transactions on Information Systems (TOIS)
  \textbf{25}(2), ~7 (2007)

\bibitem{kumar2010generalized}
Kumar, R., Vassilvitskii, S.: Generalized distances between rankings. In:
  Proceedings of the 19th international conference on World wide web. pp.
  571--580. ACM (2010)

\bibitem{lagun2014attention}
Lagun, D., Agichtein, E.: Effects of task and domain on searcher attention. In:
  Proceedings of the 37th International ACM SIGIR Conference on Research and
  Development in Information Retrieval. p. 1087–1090. SIGIR ’14,
  Association for Computing Machinery, New York, NY, USA (2014).
  \doi{10.1145/2600428.2609516}, \url{https://doi.org/10.1145/2600428.2609516}

\bibitem{lawrence1998searchwww}
Lawrence, S., Giles, C.L.: Searching the world wide web. Science
  \textbf{280}(5360),  98–100 (1998)

\bibitem{lee2005url}
Lee, S.H., Kim, S.J., Hong, S.H.: On {URL} normalization. In: International
  Conference on Computational Science and Its Applications. pp. 1076--1085.
  Springer (2005)

\bibitem{srivastava1999precision}
Leighton, H.V., Srivastava, J.: First 20 precision among world wide web search
  services (search engines). Journal of the American Society for Information
  Science  \textbf{50}(10),  870--881 (1999)

\bibitem{lewandowski2008retrieval}
Lewandowski, D.: The retrieval effectiveness of web search engines: considering
  results descriptions. Journal of Documentation  \textbf{64}(6),  915--937
  (2008)

\bibitem{zeyang2015influence}
Liu, Z., Liu, Y., Zhou, K., Zhang, M., Ma, S.: Influence of vertical result in
  web search examination. In: Proceedings of the 38th International ACM SIGIR
  Conference on Research and Development in Information Retrieval. p.
  193–202. SIGIR ’15, Association for Computing Machinery, New York, NY,
  USA (2015). \doi{10.1145/2766462.2767714},
  \url{https://doi.org/10.1145/2766462.2767714}

\bibitem{nyt}
Popper, N.: A feisty {Google} adversary tests how much people care about
  privacy. The New York Times  (Jul 2019),
  \url{https://www.nytimes.com/2019/07/15/technology/duckduckgo-private-search.html}

\bibitem{reshef2011detecting}
Reshef, D.N., Reshef, Y.A., Finucane, H.K., Grossman, S.R., McVean, G.,
  Turnbaugh, P.J., Lander, E.S., Mitzenmacher, M., Sabeti, P.C.: Detecting
  novel associations in large data sets. Science  \textbf{334}(6062),
  1518--1524 (2011)

\bibitem{ronald1998}
{Ronald}, S.: More distance functions for order-based encodings. In: 1998 IEEE
  International Conference on Evolutionary Computation Proceedings. IEEE World
  Congress on Computational Intelligence (Cat. No.98TH8360). pp. 558--563 (May
  1998). \doi{10.1109/ICEC.1998.700089}

\bibitem{schonfeld2006not}
Schonfeld, U., Bar-Yossef, Z., Keidar, I.: Do not crawl in the {DUST}:
  different {URL}s with similar text. In: Proceedings of the 15th international
  conference on World Wide Web. vol.~3, pp. 1015--1016. ACM (2006).
  \doi{10.1145/1242572.1242588}

\bibitem{etzioni1995metacrawler}
Selberg, E., Etzioni, O.: Multi-service search and comparison using the
  metacrawler. 4th International Conference on World Wide Web  (1995)

\bibitem{spink2008overlap}
Spink, A., Jansen, J., Wang, C.: Comparison of major web search engine overlap:
  2005 and 2007. AusWeb 2008: 14th Australasian World Wide Web Conference
  (2008)

\bibitem{spink2006study}
Spink, A., Jansen, B.J., Blakely, C., Koshman, S.: A study of results overlap
  and uniqueness among major web search engines. Information Processing \&
  Management  \textbf{42}(5),  1379--1391 (2006)

\bibitem{spink2006_2overlap}
Spink, A., Jansen, B.J., Blakely, C., Koshman, S.: A study of results overlap
  and uniqueness among major web search engines. Information Processing \&
  Management  \textbf{42}(5),  1379 -- 1391 (2006).
  \doi{https://doi.org/10.1016/j.ipm.2005.11.001}

\bibitem{vaughan2004new}
Vaughan, L.: New measurements for search engine evaluation proposed and tested.
  Information Processing \& Management  \textbf{40}(4),  677--691 (2004)

\bibitem{wallace2005statistical}
Wallace, C.S.: Statistical and inductive inference by minimum message length.
  Springer Science \& Business Media (2005)

\bibitem{wang2018whole}
Wang, Y., Yin, D., Jie, L., Wang, P., Yamada, M., Chang, Y., Mei, Q.:
  Optimizing whole-page presentation for web search. ACM Trans. Web
  \textbf{12}(3) (Jul 2018). \doi{10.1145/3204461},
  \url{https://doi.org/10.1145/3204461}

\bibitem{webber2010similarity}
Webber, W., Moffat, A., Zobel, J.: A similarity measure for indefinite
  rankings. ACM Transactions on Information Systems (TOIS)  \textbf{28}(4), ~20
  (2010)

\bibitem{white2009search}
White, R.W., Dumais, S.T.: Characterizing and predicting search engine
  switching behavior. 18th ACM Conference on Information and Knowledge
  Management p. 87–96 (2009)

\bibitem{winkler1990string}
Winkler, W.E.: String comparator metrics and enhanced decision rules in the
  {F}ellegi-{S}unter model of record linkage  (1990)

\bibitem{yue2010bias}
Yue, Y., Patel, R., Roehrig, H.: Beyond position bias: Examining result
  attractiveness as a source of presentation bias in clickthrough data. In:
  Proceedings of the 19th International Conference on World Wide Web. p.
  1011–1018. WWW ’10, Association for Computing Machinery, New York, NY,
  USA (2010). \doi{10.1145/1772690.1772793},
  \url{https://doi.org/10.1145/1772690.1772793}

\bibitem{zaragoza2010web}
Zaragoza, H., Cambazoglu, B.B., Baeza-Yates, R.: Web search solved?: all result
  rankings the same? In: Proceedings of the 19th ACM international conference
  on Information and knowledge management. pp. 529--538. ACM (2010)

\bibitem{zhuang2008semantic}
Zhuang, Z., Cucerzan, S.: Exploiting semantic query context to improve search
  ranking. In: Proceedings of the 2008 IEEE International Conference on
  Semantic Computing. p. 50–57. ICSC ’08, IEEE Computer Society, USA
  (2008). \doi{10.1109/ICSC.2008.8}, \url{https://doi.org/10.1109/ICSC.2008.8}

\end{thebibliography}

\end{document}